\documentclass[superscriptaddress,twocolumn,prb,showpacs,floatfix]{revtex4-1}
\usepackage[dvips]{graphicx}
\usepackage{amsmath,amsfonts,amssymb,bm,ulem}
\usepackage{graphicx}
\usepackage[usenames]{color}

\usepackage{epsf}

\begin{document}

\title{Acoustic Studies of AC Conductivity Mechanisms in $n$-GaAs/AlGaAs in the Integer and Fractional Quantum Hall Effect Regime.}

\author{I.L.~Drichko}
\affiliation{A.F. Ioffe Physical-Technical Institute of Russian
Academy of Sciences, 194021 St. Petersburg, Russia}
\author{I.Yu.~Smirnov}
\email{ivan.smirnov@mail.ioffe.ru}
\affiliation{A.F. Ioffe
Physical-Technical Institute of Russian Academy of Sciences, 194021
St. Petersburg, Russia}
\author{A.V.~Suslov}
\affiliation{National High Magnetic Field Laboratory, Tallahassee,
FL 32310, USA}
\author{D.R. Leadley}
\affiliation{Department of Physics, University of Warwick, Coventry
CV4 7AL, UK}
\date{\today}

\begin{abstract}
In case of a of the heterostructure n-GaAs/AlGaAs with sheet density
$n=2 \times 10^{11}$cm$^{-2}$ and mobility  $\mu \approx 2 \times
10^6$ cm$^2$/V$\cdot$s with integer and fractional quantum Hall
effect (IQHE and FQHE, respectively) we demonstrate the wide
applicability of acoustic methods for determining the general
conduction parameters of a two dimensional electron gas. We also
examine the mechanisms of low-temperature conductivity in the minima
of oscillations of high frequency conductivity in the IQHE and FQHE
regimes. In the magnetic field region where electrons are
delocalized, the parameters determined by the acoustic technique do
not differ from those determined by a direct current. However, the
acoustic measurements do not require Hall bars and electrical
contacts to be fabricated. In the minima of IQHE and FQHE
oscillations electrons are localized, and ac conductivity turns to
be via hopping. An analysis of the high frequency conductivity in
the QHE regime has been carried out within a "two site" model.
Furthermore, measurements of acoustoelectric effects in a tilted
magnetic field provided the dependence of the activation energy on
magnetic field in the fractional quantum Hall effect regime at
$\nu$=2/3.
\end{abstract}
 \maketitle

\section{Introduction} \label{intr}
Electrons in modulation-doped GaAs/AlGaAs have been recorded with  a
mobility of over $10^{7}$ cm$^2$/V$\cdot$s,  which has attracted
huge interest from experimentalists engaged with the physics of
two-dimensional systems. Indeed, amazing features such as the
integer and fractional quantum Hall effects~\cite{bib:1} have been
discovered, leading to two Nobel Prizes, and the properties of these
structures have been studied by numerous authors with various
methods.

In our studies, we use contactless acoustic methods to investigate
the quantum Hall effect in $n$-GaAs/AlGaAs heterostructures, grown
by molecular beam epitaxy. This technique was first employed
experimentally in Ref.~\onlinecite{bib:Wix}, with theoretical
calculations of the absorption coefficient and velocity of a surface
acoustic wave (SAW) interacting with the carriers in the two
dimensional channel made in Ref.~\onlinecite{bib:SAWcalc}. The aim
of the present paper is to demonstrate how the  kinetic
characteristics of delocalized electrons, as well as features of
low-temperature hopping conduction, can be determined by this method
using the example of  a $n$-GaAs/AlGaAs heterostructure with an
electron mobility  $\mu \approx 2\times10^{6}$ cm$^2$/V$\cdot$s and
a carrier concentration of $n=2 \times 10^{11}$cm$^{-2}$. For well
studied materials, such as $n$-GaAs/AlGaAs, the techniques required to
manufacture Hall-bar samples with ohmic contacts is well
established, which means that the material parameters can be
reliably obtained from direct current measurements. However, for a
number of new systems, where the quality of the contacts can be a
serious problem, acoustic methods provide an option to analyze the
material parameters in a contactless way, requiring only samples in
a standard rectangular form of a plate with macroscopic dimensions
of  $\approx$5 mm $\times$ 5 mm. Although there have been numerous
previous excellent studies of SAW effects in GaAs heterojunctions,
we will discuss here how the technique can be used more generally as
a substitute for contact measurements in extracting numerical
parameters of electrons/holes at the low dimensional interface.
This implementation was realized for the first time.

In a magnetic field, the SAW absorption and velocity show
Shubnikov-de Haas (SdH)-type oscillations. Although the acoustic
measurements are relative, the absolute characteristics of the
delocalized carriers (an electron gas): the carrier concentration,
mobility, conductivity at zero magnetic field, and the transport
relaxation time, can be determined in the linear regime. In the
nonlinear regime, when the ac conductivity begins to depend on the
SAW intensity, one can also determine the energy relaxation
time.~\cite{GeSiOur} Our experimental configuration, which we refer
to as "hybrid", opens up the possibility to not only explore
materials that lack an inversion center, such as
A$^{\text{III}}$B$^{\text{V}}$, but also those with an inversion
center, e.g. Ge, Si etc. Nevertheless, to obtain the numerical data
of material parameters in the hybrid experimental configuration it
is necessary to determine the gap between the sample and the lithium
niobate platelet, which is formed due to the roughness of both
surfaces. We have developed methods to determine this gap for the
first time.

Within the quantum Hall effect regime, it is usual for researchers
using acoustic methods to compare the observed experimental
dependences of the SAW attenuation on the magnetic field with the
conductivity measured with a direct current. We have shown that it
is not quite correct. In the quantum Hall minima (of the SdH
resistance oscillations and oscillations of the SAW absorption and
velocity) the conductivity has a hopping nature and in the hopping
regime the high frequency (ac) and dc conductivities have different
mechanisms: While dc-conductivity is realized by hopping of carriers
between centers that are close in \textit{energy} (variable range
hopping regime) from one end of the sample to the other, the ac
conductivity is due to hopping between two \textit{spatially} close
sites and has a complex form  $\sigma^{ac} = \sigma_1-i \sigma_2$
for the 2-dimensional case.~\cite{Efros} We have developed a
technique to determine the real and imaginary components of the
conductivity by simultaneous measurements of the magnetic field
dependences of the SAW attenuation and velocity,~\cite{bib:1stPRB}
which is particularly powerful since determination of the imaginary
part of the hopping conductivity at high frequency measurements is
usually troublesome.

When the charge carriers are delocalized, in the SAW frequency range
employed in this work, the imaginary component of conductivity is
very small and can be neglected  $\sigma_1 > \sigma_2 \approx 0$,
thus  $\sigma^{ac}=\sigma^{dc}$; however, if the carriers are
localized and conductivity is of a hopping nature the imaginary
component of conductivity can exceed the real one  $\sigma_2 >
\sigma_1$.~\cite{Efros} In the integer quantum Hall effect (IQHE)
regime, the high frequency conductivity oscillation minima
corresponds to the carriers being completely or partially localized.
In latter case, the conductivity mechanism is a mixture of hopping
via localized states and conduction of delocalized electrons through
the extended states. The ratio of the real and imaginary components
of the ac conductivity then gives a qualititative indication of  the
fraction of localized charge carriers. Such an analysis was
carried out for the first time.

The fractional quantum Hall effect (FQHE) regime has also been the
subject of detailed studies by a very large number of
researchers.~\cite{IQHE} An extensive study of the FQHE by acoustic
methods has also been performed and is reviewed in
Ref.~\onlinecite{Willett}. However, much of this survey focused on
an analysis of  the magnetic field region close to filling factor
$\nu$=1/2, i.e., the one used to argue theoretically the existence
of composite fermions (CFs).~\cite{CFtheor} The dispersion of
excitations in the fractional Hall effect regime, using
high-frequency SAWs, was studied in Ref.~\onlinecite{Kukush}.

In our study, we investigated the ac conductivity by acoustic
methods at filling factor  $\nu$=2/3, which falls at $B\approx$12 T
in our sample. The FQHE at  $\nu$=2/3 is of great interest, because
in the CF model this state consists of two filled CF Landau levels
(LLs) and so can be either unpolarized, at low Zeeman energy, or
fully spin polarized at higher Zeeman energy. The spin transition
between these states has been investigated by tilting the sample in
a  magnetic field to increase the total magnetic field (setting the
Zeeman energy) for a given perpendicular component (that determines
the filling factor).~\cite{Eisen,2x3f,2x3h,Clark} As an alternative
the Zeeman energy can be reduced, all the way to zero, by using
hydrostatic pressure to alter the band structure.~\cite{14a, 14b}
However, in many of the tilted magnetic field
studies,~\cite{Schulze,Haug,Boebinger,Khrapai} the carrier
concentration in the sample was too high to directly observe the
paramagnetic-ferromagnetic transition at  $\nu$=2/3  associated with
the crossing of the CF Landau levels, as is also the case for our
sample. Instead these studies measured the magnetic field dependence
of the activation energy on the ferromagnetic side. It is assumed
that, based on a simple model of composite fermions, the energy gap
in the FQHE regime should be determined by the Coulomb interaction
in the form $\Delta E \propto e^2/\varepsilon l_B$, where
$l_B=\sqrt{\hbar c / e B}$ is the magnetic length, and therefore
$\Delta E \propto \sqrt{B}$.

However, this dependence was not observed in any of the above tilted
field experiments. Furthermore, in the pressure experiments it was
observed that the CF energy gap scaled like  $\sqrt{B}^*$, where
$B^*=B-2(h/e)n$ is the effective magnetic field in which the CF
moves,~\cite{14a} as might be expected given the fractal nature of
the QHE.~\cite{18a}  Therefore, it is useful to also determine the
dependence of $\Delta E (B)$ by acoustic contactless methods. A
study of the FQHE using acoustic methods at $\nu$=2/3 was
implemented in Ref.~\onlinecite{Dini}, motivated by the conclusion
from dc measurements  that domains are formed in the vicinity of the
spin transition, to determine the width of these domain walls.
However, in the frequency range 100 MHz - 2 GHz no effects on the
SAW propagation have been observed in the spin transition region.

In addition to the above mentioned aims, the purpose of our present
work is to determine the conductivity mechanism for electrons in
FQHE regime ($\nu$=2/3) in a fully spin-polarized - ferromagnetic
region and to determine the dependence of the energy gap of this
state on the magnetic field in one more way, by measuring the
high-frequency conductivity acoustically.

\section{Experimental method} \label{exp}

The complex high frequency conductivity $\sigma^{\text{ac}} \equiv
\sigma_1 -i\sigma_2$ was investigated in a $n$-GaAs/AlGaAs sample,
over the temperature range 0.3-4.2 K and with a perpendicular and
tilted magnetic field  of up to 18 T. The sample had a carrier
concentration of $n\approx$2$\times$10$^{11}$\,cm$^{-2}$ and
mobility $\mu \approx 2\times10^{6}$ cm$^2$/V$\cdot$s, as obtained
from dc transport measurements in the dark. This investigation used
acoustic methods to measure the absorption of a  SAW $\Gamma$  and
it's velocity shift in magnetic field - $\Delta v(B)/v (0)$. The SAW
frequency was varied from 30 to 254 MHz.
\begin{figure}[ht]
\centerline{
\includegraphics[width=8cm]{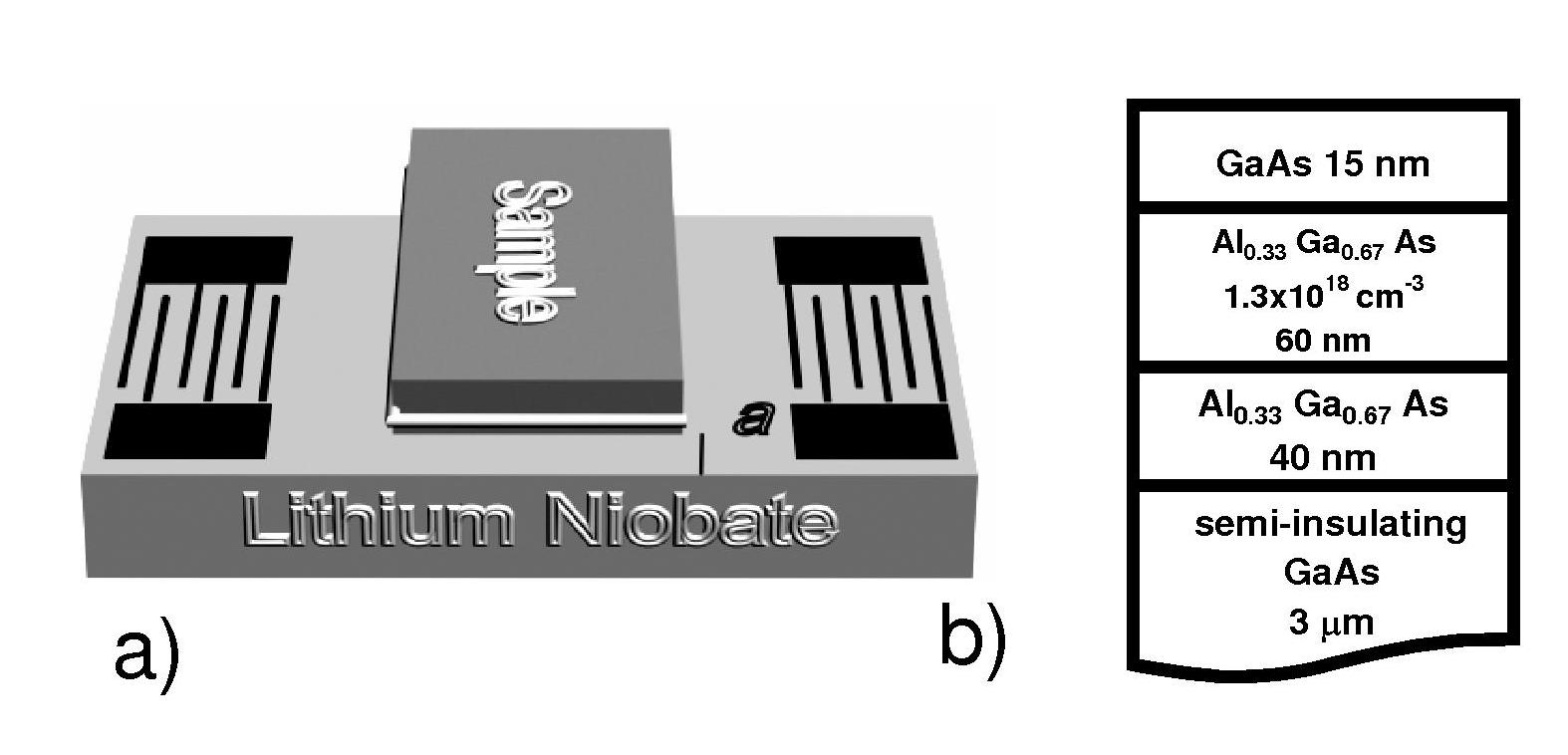}
} \caption{Sketches of the (a) acoustic experiment setup and (b) sample.
 \label{fig:sample}}
\end{figure}

Acoustic methods, based on the use of a surface acoustic wave,
provide an opportunity to work with samples of a square or
rectangular shape without fabricating electrical contacts, which can
otherwise plague the interpretation of dc transport measurements at
low temperature. For this purpose the "hybrid" method was used,
whereby the SAW, excited by interdigital transducers, propagates
along the surface of the piezodielectric lithium niobate, and the
structure to be studied is slightly pressed onto the surface of the
LiNbO$_3$ substrate by means of springs (shown in
Fig.~\ref{fig:sample} a). The acoustic deformation wave, propagating
along the surface of the LiNbO$_3$, is accompanied by a wave of
electric field with the same frequency. This electric field
penetrates into the 2D channel of the GaAs sample, exciting
high-frequency currents, which in turn lead to absorption of the
wave energy. Interaction of the SAW electric field with the
electrons in a quantum well also results in a shift of the SAW
velocity. In this experimental configuration, the strain is not
transmitted to the studied system due to an effective clearance
between the sample and niobate.~\cite{bib:1stPRB}

From the experimentally measured values of the SAW absorption
$\Delta \Gamma = \Gamma(B) - \Gamma(0)$ and the relative change of
the SAW velocity $\Delta v/ v = (v (B)-v (0)) / v (0)$ in a magnetic
field one can calculate the real $\sigma_{1}$ and imaginary
$\sigma_{2}$ components of high-frequency conductivity using the
equations (1) and (2).~\cite{bib:SAWcalc} Since $\Gamma(B=0) \ll
\Gamma(B)$:
\begin{eqnarray}
  \label{eq:G}
&&\Gamma=8.68\frac{K^2}{2}qA  \times  \,   \nonumber \\
&& \times   \frac{4\pi\sigma_1t(q)/\varepsilon_sv}
  {[1+4\pi\sigma_2t(q)/\varepsilon_sv]^2+[4\pi\sigma_1t(q)/\varepsilon_sv]^2}, \frac{\text{dB}}{\text{cm}}  \,   \\
&&\text{where }  A = 8b(q)(\varepsilon_1 +\varepsilon_0)
\varepsilon_0^2 \varepsilon_s
\exp [-2q(a+d)],\text{ and}  \,   \nonumber \\
\label{eq:V}
&&\frac{\Delta v}{v}=\frac{K^2}{2}A \times  \,   \nonumber \\
&&\times   \frac{1+4\pi\sigma_2t(q)/\varepsilon_sv}
  {[1+4\pi\sigma_2t(q)/\varepsilon_sv]^2+[4\pi\sigma_1t(q)/\varepsilon_sv]^2},
\end{eqnarray}
where $K^2$ is the electro-mechanic coupling constant for lithium
niobate (Y-cut in our experiments), $q$ and $v$ are the SAW wave
vector and velocity in LiNbO$_3$, respectively; $a$ is the gap
between the piezoelectric plate and the sample, $d$ is the distance
between the sample surface and the 2DEG layer; $\varepsilon_1$,
$\varepsilon_0$ and $\varepsilon_s$ are the dielectric constants of
LiNbO$_3$, of
 vacuum, and of the semiconductor, respectively; $b(q)$ and $t(q)$ are complex
functions of $a$, $d$, $\varepsilon_1$, $\varepsilon_0$ and
$\varepsilon_s$ given in Ref.~\onlinecite{bib:SAWcalc}. Using these
equations one may obtain $\sigma_1$ and $\sigma_2$ provided that the
values of $a$ and $d$ are known. The technological process of sample
production determines d, and a is a fitting parameter which is
determined  in an appropriate way for the effect being studied (see
below). This value depends both on the procedure of mounting the
sample  on the lithium niobate surface as well as on the quality of
sample and niobate surfaces.

The  modulation doped GaAs heterostructure (shown in
Fig.~\ref{fig:sample} b) was grown by molecular beam epitaxy at
Philips Redhill~\cite{19a} starting from a semi-insulating GaAs
substrate with 3$\mu$m of epitaxial GaAs forming the conducting
channel where the two dimensional electron gas (2DEG) is located,
followed by  a 40 nm undoped  Al$_{0.33}$Ga$_{0.67}$As spacer layer
followed by a 60 nm supply layer of  Al$_{0.33}$Ga$_{0.67}$As doped
with a Si  donor concentration at $1.3 \times 10^{18}$ cm$^{-3}$,
and finally a 15 nm GaAs cap.

\section{Experimental results and discussion} \label{exp and res}

\subsection{Linear regime} \label{Linear regime}

\begin{figure}[ht]
\centerline{
\includegraphics[width=8.5cm]{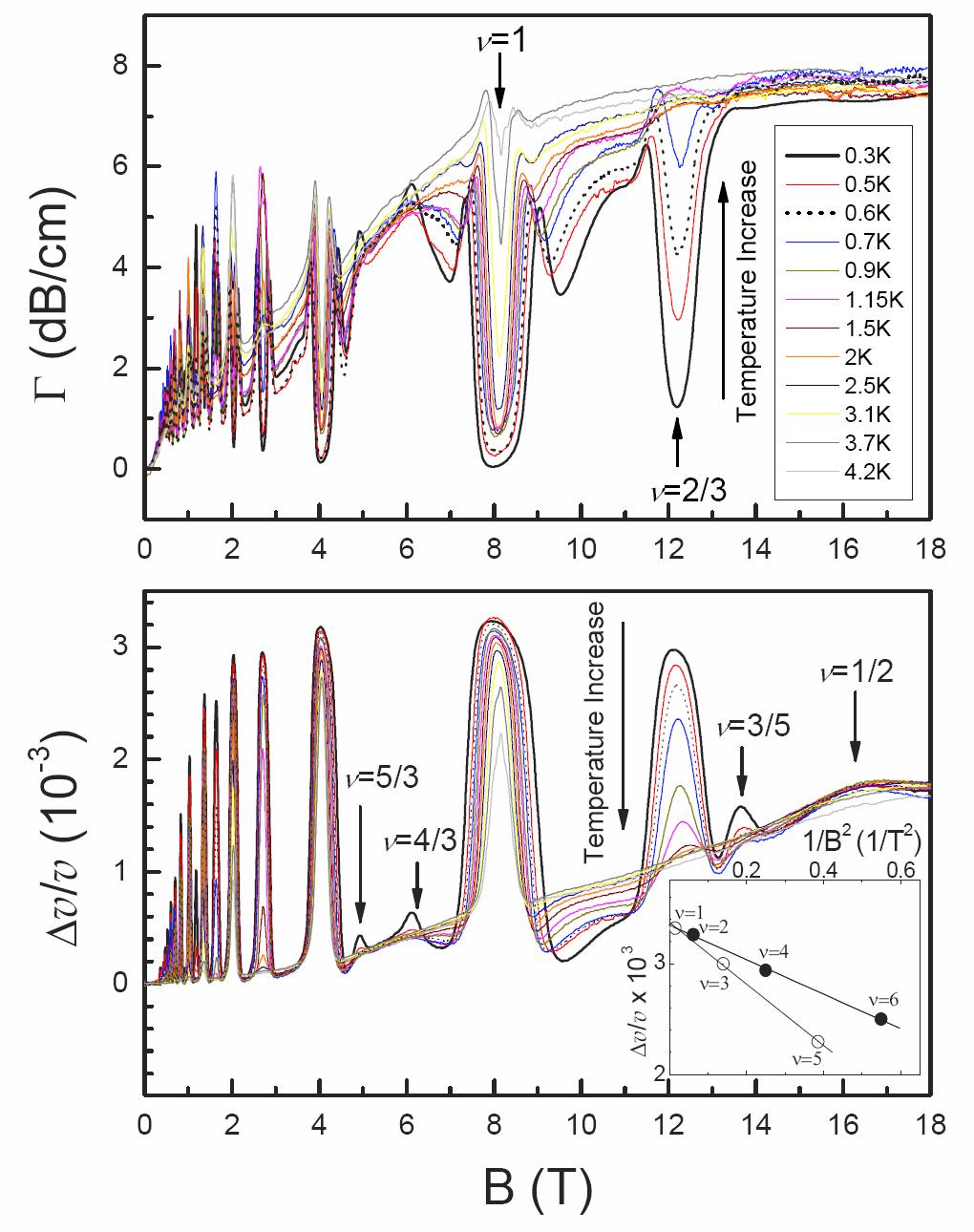}
} \caption{(Color online) Magnetic field dependences of
the ultrasound attenuation $\Gamma$ and velocity shift
$\Delta
  v/v(0)$ for different temperatures; $v(0)\approx$3.5$\times$10$^{5}$~cm/s is the zero-field SAW velocity; $f=30$ MHz.
  Inset: the dependence of the maxima $\Delta v / v$ on $1/B^2$
for odd and even filling factors; $f=30$ MHz, $T$=0.3 K; lines
are results of the linear fit.
 \label{fig:GVgaas}}
\end{figure}

We will first discuss the results obtained in the linear regime
where the response does not depend on the input SAW power.  For this
section the output from the RF generator was kept below 1 $\mu$W .
Fig.~\ref{fig:GVgaas} shows the magnetic field dependence of the
attenuation $\Gamma$ and the SAW velocity change  $\Delta v/v(0)$ at
different temperatures from 0.3 K to 4.2 K using a SAW frequency of
$f$=30 MHz.

One can see that magnetic field induces oscillations of the
absorption coefficient and the velocity change, analogous to the
SdH oscillations seen in $\rho_{xx}$ when
using a direct current. These extrema vary with temperature much
like SdH oscillations with the odd integer features disappearing faster with
increasing temperature than the even integer features.  In addition
FQHE features can be seen near to filling factor 2/3 and weaker ones
5/3, 4/3 and 3/5 as well as the temperature independent point at
1/2.

According to Eq.2, the value of  $\Delta v / v$ should increase as
$1/B^2$ and saturate at the value $K^2 A / 2$ in a strong magnetic
field. Such behavior is seen in Fig.~\ref{fig:GVgaas}, where the
maximum value at different  $\nu$ does tend to saturate.
Furthermore, if we construct the dependence (see inset of
Fig.~\ref{fig:GVgaas}) of the maxima $\Delta v / v$ on $1/B^2$
separately for odd filling factors 1,3,5... and even filling factors
2,4,6... we obtain two straight lines, with different slopes, which
intersect at $1/B^2$=0. This intersection is equal to the saturation
value, from which we can deduce the empirical parameter $a$,
$a\approx (0.8-1.0) \times 10^{-4}$ cm.

Having found $a$, we can now calculate the conductivity  $\sigma_1$
and $\sigma_2$ from the measured $\Gamma$ and $\Delta v/v(0)$ by
using equations (1) and (2).

\begin{figure}[ht]
\centerline{
\includegraphics[width=8.5cm]{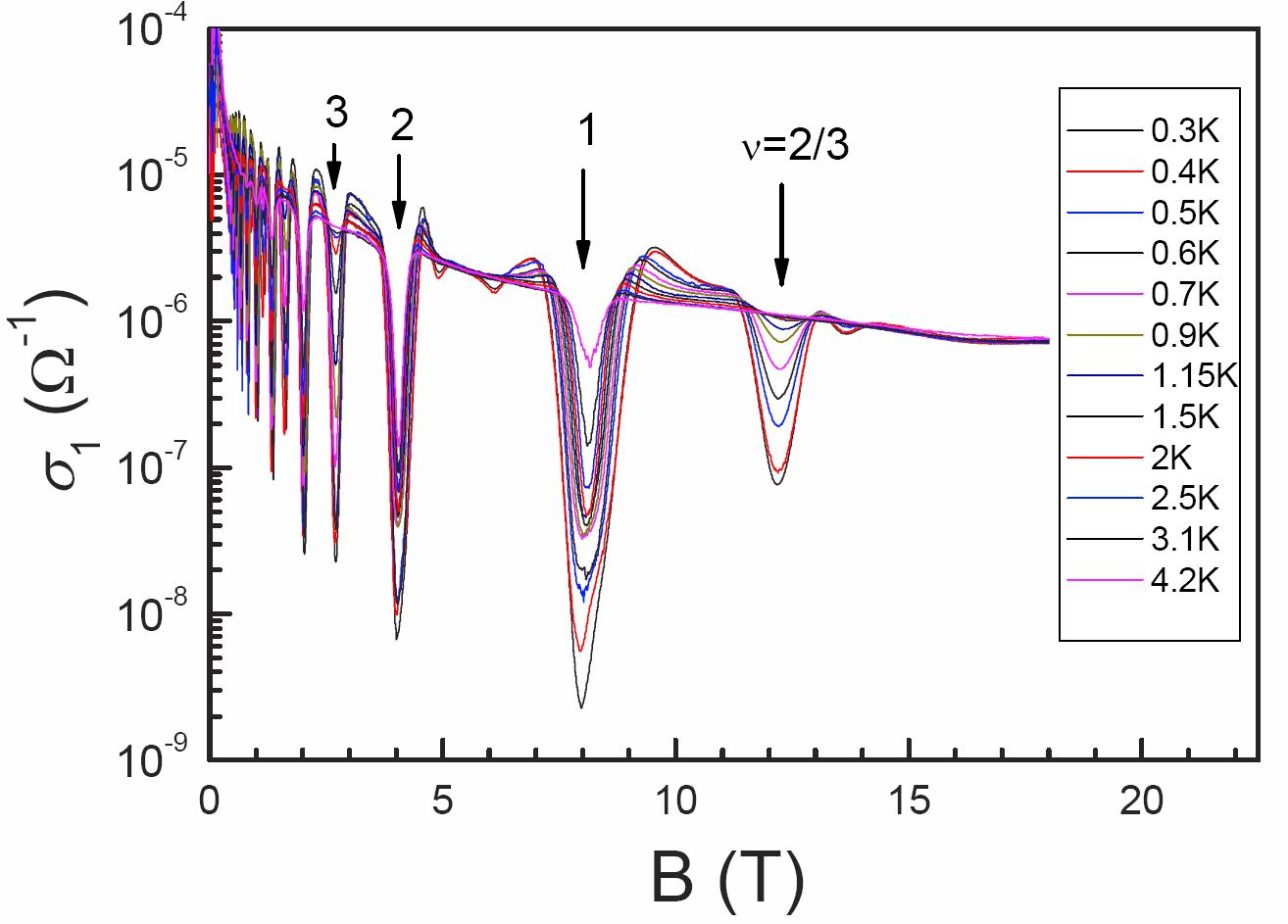}
} \caption{(Color online) Magnetic field dependence of the real part
of the ac conductivity $\sigma_1$ at temperatures from 0.3 K to 4.2 K; $f =$30 MHz.
 \label{FigS1Bgaas}}
\end{figure}

The dependence of the real part of the ac conductivity in the
two-dimensional channel $\sigma_1$ on the magnetic field, derived
from the curves in Fig.~\ref{fig:GVgaas} is represented in
Fig.~\ref{FigS1Bgaas}, where it can be seen that $\sigma_1$ changes
by 5 orders of magnitude, from 10$^{-4}$ at low field to
$\approx$10$^{-9}$ Ohm$^{-1}$ in the minima of the SdH oscillation.
Naturally, the
high conductivity in weak magnetic fields and low conductivity in
the oscillation minima is implemented by different mechanisms.
Therefore, we will consider the mechanisms of conductivity for the
different regions separately and show the feasibility of using
acoustic methods to determine the parameters of the material under
these different conditions.

\subsubsection{Regime of the free charge carriers, $B<$1 T}
\label{free}

\begin{figure}[ht]
\centerline{
\includegraphics[width=8.5cm]{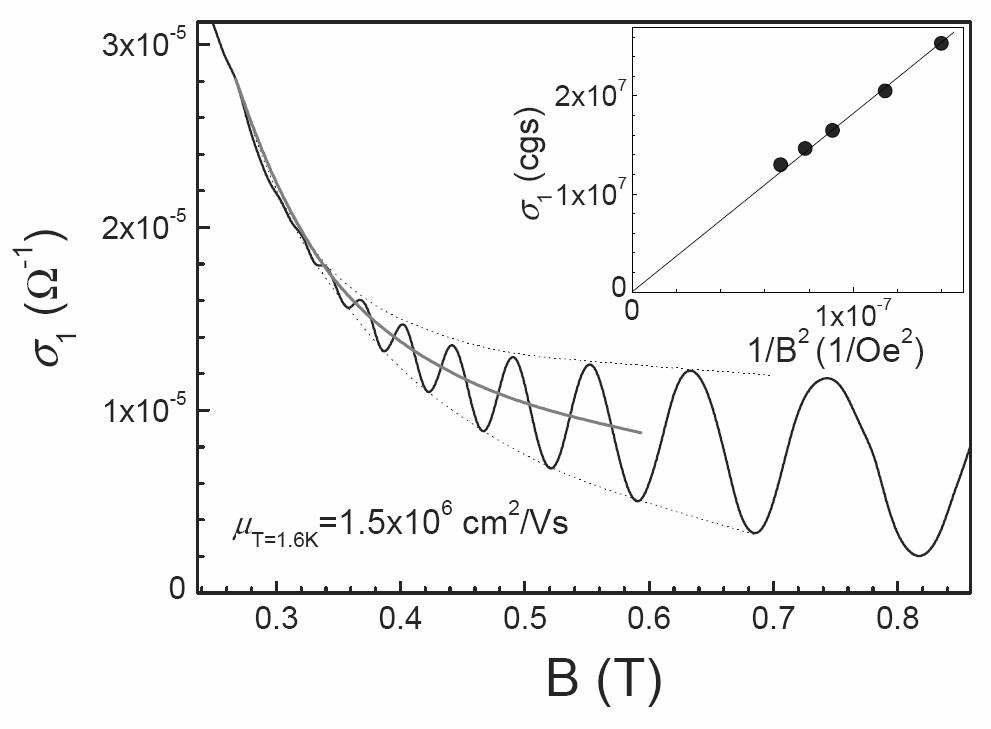}
} \caption{Magnetic field dependence of the real part
of the ac conductivity $\sigma_1$ in low fields. The dotted lines are the oscillations envelope;
the solid curve is the mean of envelope. Inset: Nonoscillating term of the conductivity $\sigma_{xx}^*$ vs 1/$B^2$;
$f$=30 MHz, $T$=1.6 K; line is a result of the linear fit.
 \label{FigMobil}}
\end{figure}

Measurements of the SAW absorption coefficient $\Gamma$ and velocity
shift $\Delta v(B)/v(0)$ are relative, since they are usually
carried out by changing  some external impact (magnetic or electric
field, irradiation, etc.). Hence, it is impossible to directly
determine the absolute values of conductivity, or the mobility of
charge carriers at $B$=0 from these measurements. However, these
parameters can be found in high mobility samples, which exhibit many
SdH oscillations in relatively small magnetic fields, where the
charge carriers are delocalized and the dc conductivity
$\sigma^{dc}=\sigma_1 \gg \sigma_2 \approx 0$. In accordance with
Ref.~\onlinecite{Ando}:
\begin{eqnarray}
  \label{eq:01x}
&& \sigma_{xx}=\sigma_{xx}^*+\sigma_{xx}^{osc} \,   \nonumber \\
&& \sigma_{xx}^*=\frac{\sigma_{0}}{1+\omega_c^2\tau_0^2}=
  \frac{enc^2}{\mu B^2}, \text{ provided } (\omega_c^2\tau_0^2)\gg 1
\end{eqnarray}
where $\sigma_{0}$ is the zero magnetic field conductivity,
$\omega_c=eB/m^*c$ is the cyclotron frequency, $\tau_0$ is the
transport relaxation time; $c$=3$\times 10^{10}$ cm/s, $e$ is the
electron charge, $n$ is the sheet density found from the
oscillations period. Fig.~\ref{FigMobil} is an expanded view of the
real part of conductivity  $\sigma_1$, shown in
Fig.~\ref{FigS1Bgaas}, for magnetic fields below 1 T. By drawing the
oscillation envelope, it is possible to determine $\sigma_{xx}^*$ as
the average value. Further, by plotting $\sigma_{xx}^*$ against
1/$B^2$, as illustrated in the inset to Fig.~\ref{FigMobil} for
$T$=1.6 K and $f$=30 MHz, one can determine the mobility  from the
slope of this line as $\mu= 1.5\times 10^{6}$ cm$^2$/V$\cdot$s, with
$\sigma_{xx} (B=0)$ = 4.8$\times$10$^{-2}$ Ohm$^{-1}$, which are
consistent with the results of dc transport measurements on Hall
bars taken from the same original wafer. It should be noted that
this analysis is only valid in the case where conduction is due to
delocalized electrons and the real part of the ac conductivity is
equal to the dc value $\sigma_1=\sigma_{xx}^{dc}$ and does not
depend on the frequency of SAW, with $\sigma_2 \approx$0.

Detailed analysis of the temperature evolution of the conductivity
in low fields allows the full  set of carrier parameters such as
effective mass, Dingle temperature ($T_D \approx 0.6$ K) and quantum
relaxation time ($\tau_q \approx 2.2 \times 10^{-12}$ s) to be
determined. As these methods are well known from dc transport
analysis we will not present the details here, but merely mention
the results.

\subsubsection{Integer Quantum Hall Effect Regime} \label{IQHEreg}

Let us consider the conductivity mechanisms realized in the IQHE
minima with odd filling factor  $\nu$= 1, 3, 5, 7, 9. In this case,
the charge carriers, electrons, are localized and the high-frequency
conductivity has a complex form  $\sigma^{ac}=\sigma_1-i\sigma_2$
which begins to depend on the SAW frequency. In contrast to the low
field case, we now have $\sigma_2 > \sigma_1$ and $\sigma^{ac} \neq
\sigma_{xx}^{dc}$. This condition arises because the mechanisms of
hopping conduction  for localized charge carriers are different in
ac and dc.~\cite{Efros} Hence, the existing practice of substituting
the dc conductivity in a formula like (1) is not valid when
localization of carriers takes place in a magnetic field.

\begin{figure}[ht]
\centerline{
\includegraphics[width=8.5cm]{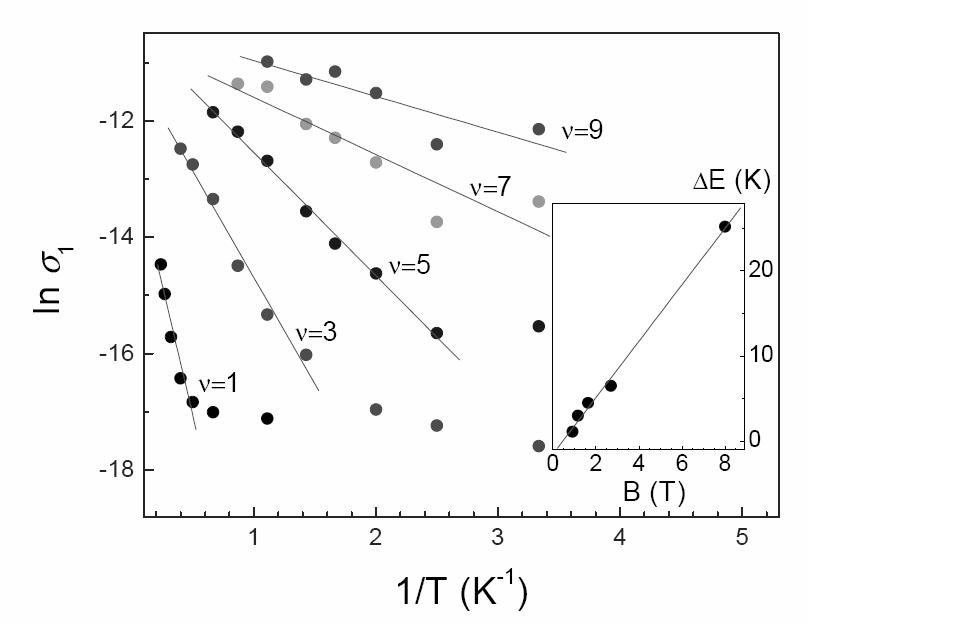}
} \caption{Arrhenius plot of the real component of the ac conductivity   $\sigma_1$ at
different odd filling factors. Inset: Activation energy in magnetic field for odd filling factors.
Lines are results of the linear fit.
 \label{FigS1T}}
\end{figure}

Fig.~\ref{FigS1T} shows how the conductivity for different odd
$\nu$, calculated by eqs. (1) and (2), varies with the inverse
temperature 1/$T$.

Figure~\ref{FigS1T} demonstrates that at relatively high
temperatures all curves show linear dependences on this scale,
corresponding to an Arrhenius law  $\sigma_1 \propto \exp(\Delta
E/2k_BT)$, allowing the activation energy $\Delta E$ to be
determined. Despite the fact that the change in conductivity is not
too large the resulting activation energy $\Delta E \gg k_BT$, so we
can use these values in what follows. An activation-like temperature
dependence of conductivity is usually associated with electron
nearest neighbor hopping in the carrier localization regime. At the
low temperature extreme of Fig.~\ref{FigS1T} the curves flatten off,
which is conventionally attributed to a change of the hopping
mechanism. This flattening occurs for all filling factors so is not
just an experimental artifact of $\sigma_1$ reaching the minimum
measurable level. In a high-frequency electric field, the electron
hops are between two closely spaced sites (the "two-site"
model~\cite{Pollak}), while at dc variable range hopping dominates.

At odd filling factors the activation energy is associated with the
formation of spin gaps, so  $\Delta E \propto g^* \mu_B B$, (where
g* is the effective g-factor, $\mu_B$ is the Bohr magneton). Thus if
one plots $\Delta E$ against $B$, as in the inset of
Fig.~\ref{FigS1T}, it is possible to determine the g-factor. The
slope of the linear dependence gives $g^*$=4.9.

\begin{figure}[ht]
\centerline{
\includegraphics[width=8.5cm]{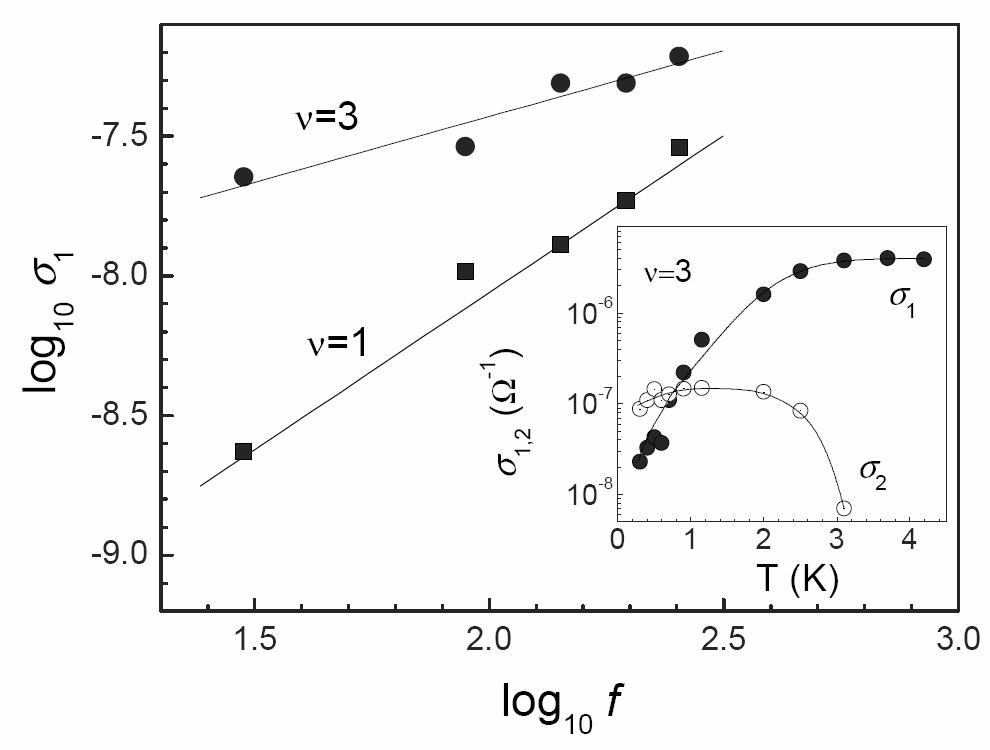}
} \caption{Dependence of the real component of the ac conductivity $\sigma_1$ on frequency at $T$=0.3 K
at filling factors $\nu$=1 and 3.
Inset: Dependences of the real $\sigma_1$ and imaginary $\sigma_2$ components of ac conductivity on
temperature at filling factor $\nu$=3; $f$=30 MHz. The lines are guides to the eye.\label{S1f}}
\end{figure}

Now let us analyze the frequency dependence of the conductivity.
Figure~\ref{S1f} shows the real part of conductivity as a function
of SAW frequency on log-log plot to show the power law of this
dependence. For $\nu$=1 $\sigma_1 \propto \omega^{1.1}$, and at
$\nu$=3 $\sigma_1 \propto \omega^{0.5}$. Thus, for $\nu$=1 $\sigma_1
\propto \omega T^0$. Such a dependence of ac conductivity on
temperature and frequency is characteristic of hopping conductivity.
This is described by the "two-site" model  provided  $\omega \tau
<1$, where $\tau$  is the time  for carriers to hop between energy
levels within the pair and $\omega=2\pi f$ is the SAW frequency. The
experimental data show that at high magnetic fields this dependence
is well satisfied. The weaker frequency dependence at $\nu$=3 could
be explained by assuming that at lower magnetic fields the
conduction mechanism is not purely hopping, as not all electrons are
localized.

If the complex ac conductivity $\sigma^{ac}=\sigma_1 - i\sigma_2$
has a hopping nature then the imaginary part of the conductivity
should exceed the real one, i.e.: $\sigma_2 >
\sigma_1$.~\cite{Efros} This theoretical condition is met as
demonstrated in the inset of the Figure~\ref{S1f}.

Inset of the Figure~\ref{S1f} shows that at very low temperatures,
where carriers are localized and "two-site" hopping conductivity is
realized, $\sigma_2
> \sigma_1$.
However, with increasing temperature when the transition to
activated conduction occurs $\sigma_1$ becomes greater than
$\sigma_2$. As the temperature increases the amount of delocalized
carriers also increases,  driving $\sigma_2$ to decrease. Finally,
$\sigma_2$ becomes vanishing for fully delocalized carriers. It
should be noted that in samples with a mobility of $2 \times 10^5$
cm$^2$/Vs this ac hopping conduction in the IQHE minima is shunted
by a high frequency hopping  conductivity via levels of
DX-centers.~\cite{DX}

\subsubsection{Regime of the Fractional Quantum Hall Effect at
$\nu=2/3$ } \label{2x3X}

\begin{figure}[ht]
\centerline{
\includegraphics[width=8.5cm]{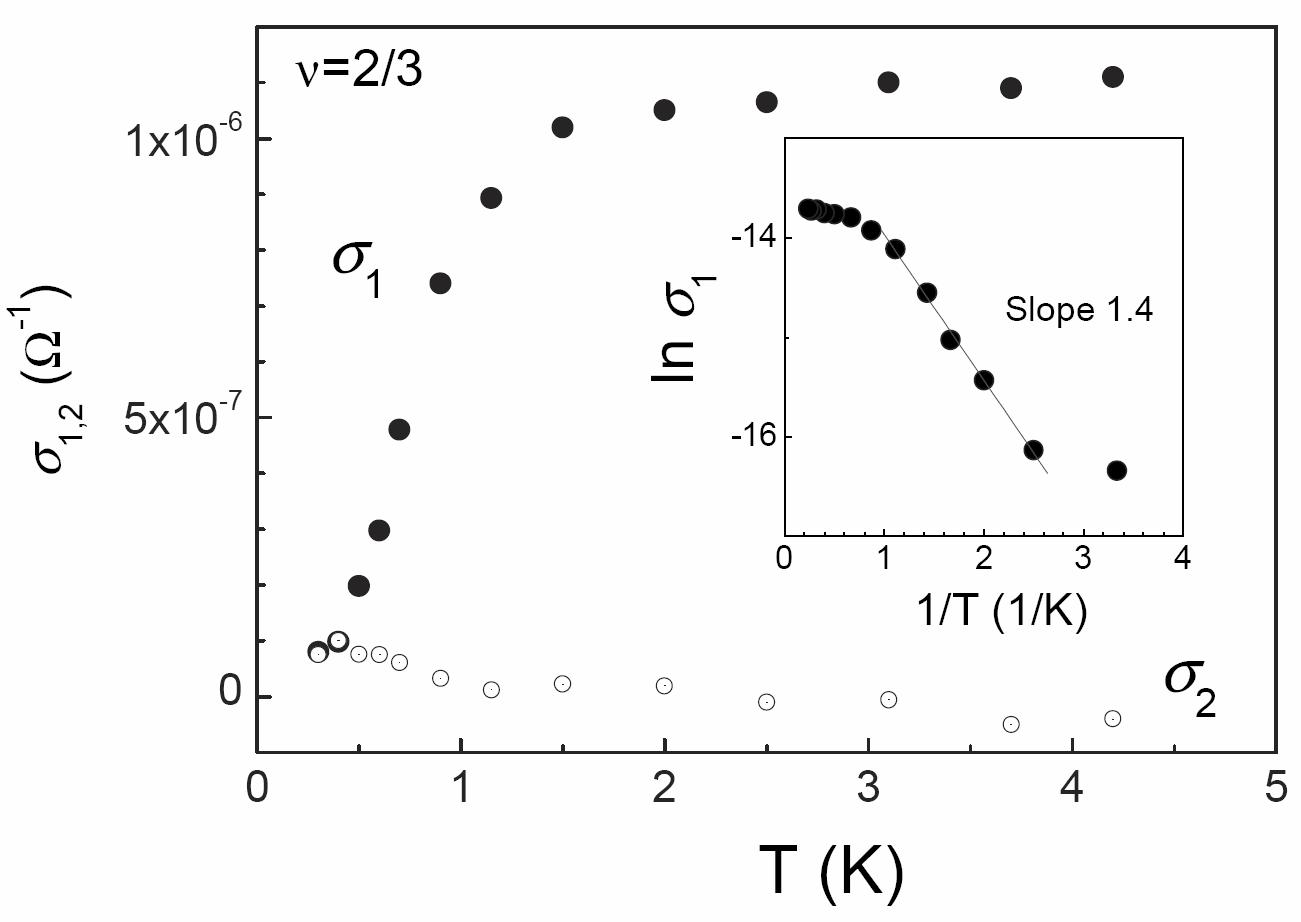}
} \caption{Dependence of the real $\sigma_1$ and imaginary $\sigma_2$ components of the ac
conductivity on temperature for filling factor $\nu$=2/3.
Inset: Arrhenius plot for $\nu$=2/3; $f$=30 MHz. \label{S12nu2x3}}
\end{figure}

Acoustic measurements can also be used to draw conclusions about the
conductivity mechanism in the fractional quantum Hall effect regime,
by considering the temperature and frequency dependences of the real
and imaginary parts of ac conductivity derived from equations (1)
and (2). The temperature dependence of  both these components for
filling factor  $\nu$=2/3 are presented in Figure~\ref{S12nu2x3},
which shows that for 0.3 K$< T <$0.5 K  $\sigma_1$ and $\sigma_2$
are the same size whereas at higher temperatures $\sigma_1 \gg
\sigma_2 \rightarrow 0$. The inset graph illustrates the dependence
of  $\sigma_1$ on inverse temperature, and provides evidence that
activated conductivity takes place in the temperature range from 0.4
K to 1 K. As previously there is a flattening at the low temperature
extreme $T<$0.4 K.

\begin{figure}[ht]
\centerline{
\includegraphics[width=8cm]{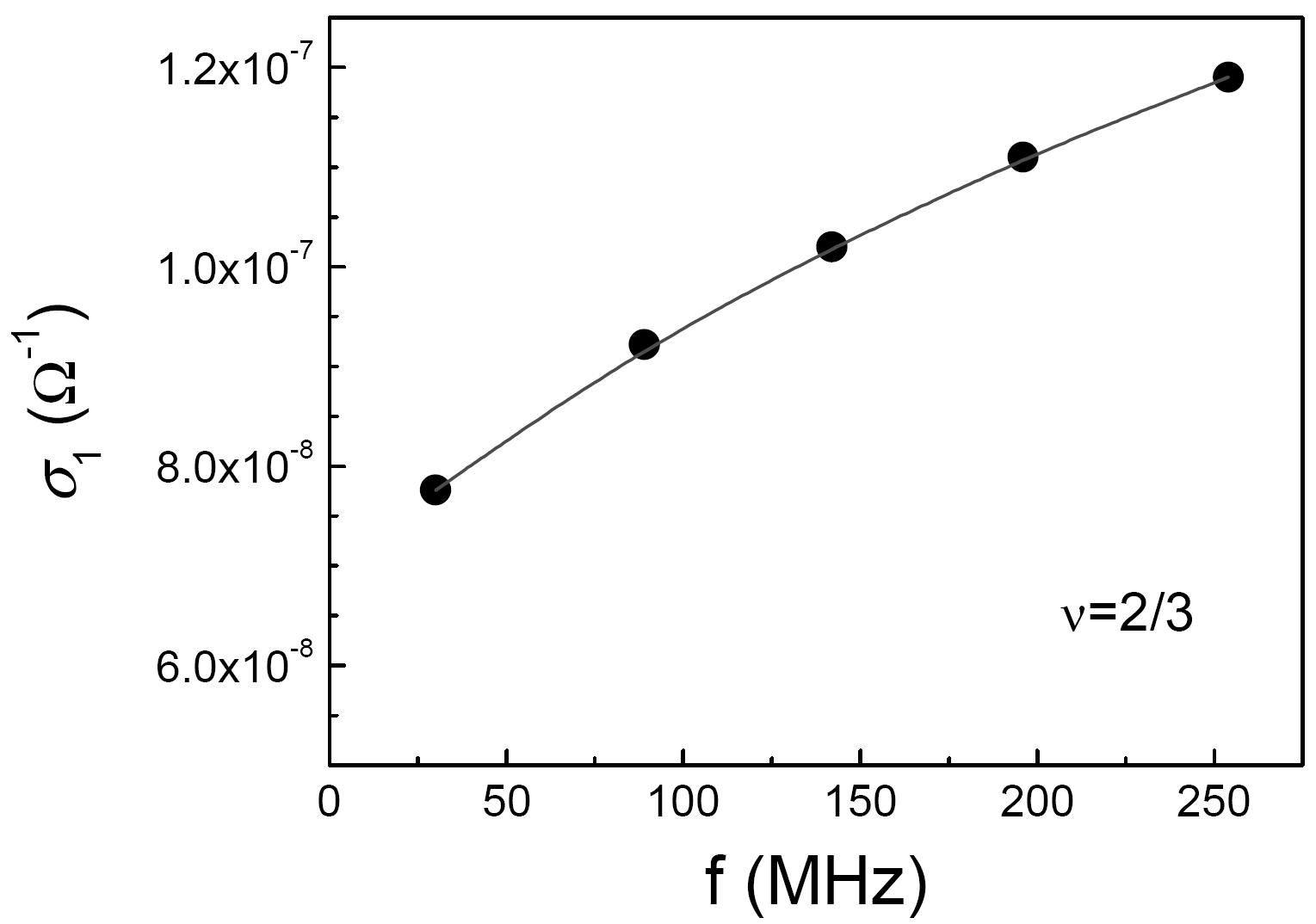}
} \caption{Frequency dependence of the real component of the ac
conductivity for filling factor $\nu$=2/3; $T$=0.3 K. The line is guide to the eye.\label{S1fnu2x3gaas}}
\end{figure}

By varying the SAW frequency, a weak frequency dependence of
$\sigma_1$ can also be seen at the lowest temperatures measured,
which is illustrated in Figure~\ref{S1fnu2x3gaas}.  These facts
indicate that even at $T$=0.3 K, the amount of localized electrons
is still small, although the hopping mechanism starts to have an
effect on the conductivity.

\subsection{Nonlinear regime} \label{nonlin}

By using larger amplitude surface acoustic waves the non-linear
regime can be entered where the response of the 2DEG depends on the
SAW intensity.  The absorption coefficient $\Gamma$  and velocity
shift $\Delta v / v$ were measured at the base temperature of
$T$=0.3 K as a function of the SAW intensity, which was varied from
0.5$\mu$W to 5 mW (as measured at the output from the RF generator).
We note that the acoustic measurements are carried out in a pulsed
mode, so the amount of power being supplied to the 2DEG is neglible
and does not simply cause an increase in temperature of the
material. This is verified in the experiments, where a sensor
mounted on the sample detected no change in sample temperature with
SAW application.
\begin{figure}[h]
\centerline{
\includegraphics[width=8.5cm]{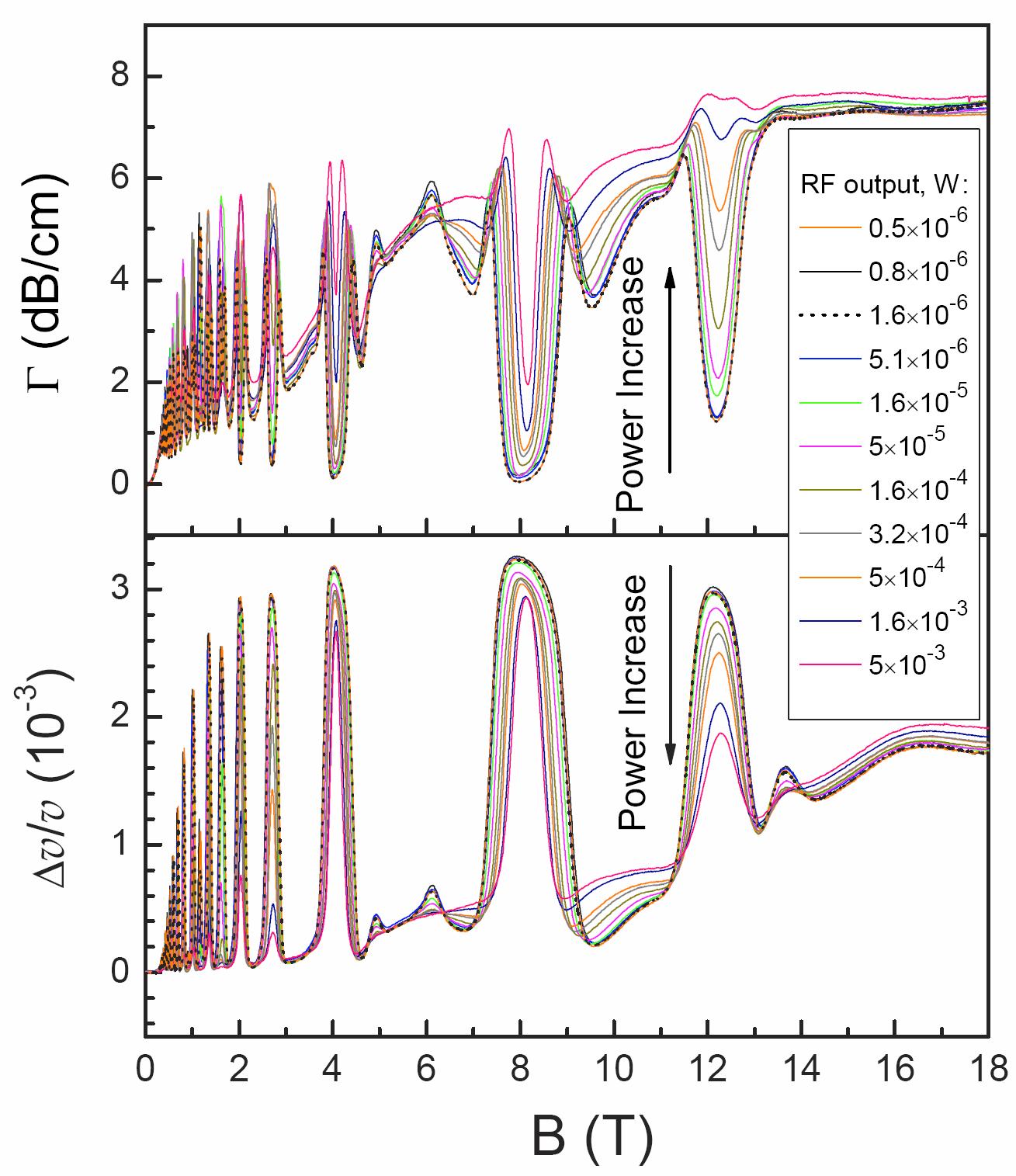}
} \caption{(Color online) Power evolution of
the ultrasound attenuation $\Gamma$ and relative change of velocity $\Delta v/v(0)$ dependences on the magnetic field;
$f$ = 30 MHz, $T$=0.3 K. \label{GamVelPower}}
\end{figure}

The attenuation $\Gamma$ and velocity shift $\Delta v/v$ in magnetic
fields up to 18 T are displayed in Figure~\ref{GamVelPower} at
different SAW powers. It is seen that increasing the SAW intensity
results in an increase of $\Gamma$ in all the minima and a decrease
of in the maxima of $\Delta v / v$. The following
Figure~\ref{S1Power} shows the real part of conductivity $\sigma_1$,
calculated using the formulas (1 and 2), as a function of the
magnetic field at different SAW intensities. Here it is seen that
with increasing intensity the ac conductivity in the oscillations
minima increases.
\begin{figure}[ht]
\centerline{
\includegraphics[width=8.5cm]{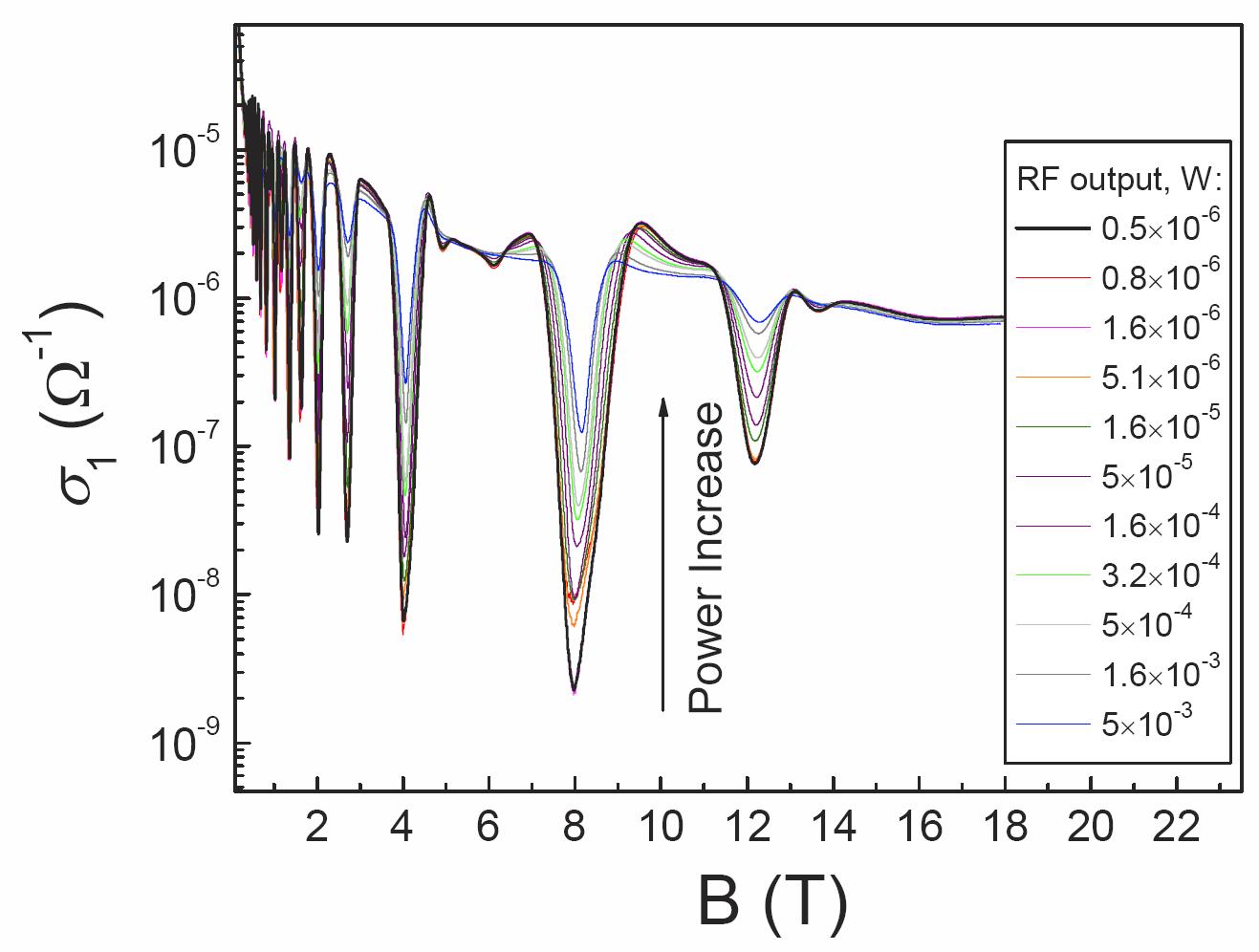}
} \caption{(Color online) Dependence of the real part of the ac conductivity  on magnetic
field at different SAW intensities; $T$=0.3 K. \label{S1Power}}
\end{figure}

The mechanisms of non-linearities that arise with increasing SAW
intensity depend on whether the electrons in the 2D channel are
localized or delocalized. Therefore we will analyze  each
conductivity regime separately.

\subsubsection{Region of delocalized carriers, $B <$ 1 T} \label{nonlinX}

The high electric field of the intense SAW field leads to heating of
the free charge carriers (Ref.~\onlinecite{HeatSAW} and references
therein). In this case, one can introduce the concept of an electron
temperature $T_e$, which is above the lattice temperature $T_p$.
Then  $T_e$ can be determined by comparing the dependence of
$\Gamma$ on  lattice temperature measured in linear regime with the
dependence of $\Gamma$ on the SAW intensity measured at the lowest
temperature.

The condition for introducing a carrier temperature is

\begin{eqnarray}
\tau_p << \tau_{ee} << \tau_{\varepsilon}, \label{taurel}
\end{eqnarray}
where $\tau_p$, $\tau_{ee}$, and $\tau_{\varepsilon}$ are the
momentum, electron-electron, and energy relaxation times,
respectively.

The electron-electron relaxation time is given by
\begin{eqnarray}
\frac{\hbar}{\tau_{ee}}=\frac{k_B T \rho e^2}{h} \ln \frac{h}{2
e^2 \rho}, \rho=1/\sigma. \label{tauee}
\end{eqnarray}
which leads to a value of  $\tau_{ee}=5 \times 10^{-9}$ s.  The
momentum relaxation time has been found from the mobility at $T$=1.6
K $\tau_p=6 \times 10^{-11} ~\text{s}$.

As regards the energy relaxation time, its calculation will be
discussed later.

To determine the absolute energy loss rate $Q=e \mu E^2$ per
electron, it is necessary to know the electric field which accompanies
the SAW. This value is determined by the formula ~\cite{HeatSAW} (if
$\sigma_2$ = 0):
\begin{eqnarray}
\ |E|^2=K^2\frac{32\pi}{v}
\frac{(\varepsilon_1+\varepsilon_0)zqe^{(-2q(a+d))}}
{[1+(\frac{4\pi \sigma_{1}(\omega)}{\varepsilon_s
v}t(q))^2]}W,\label{eq2x}
\end{eqnarray}
\begin{eqnarray}
z(q)= [(\varepsilon_1+\varepsilon_0)(\varepsilon_s+\varepsilon_0)
-(\varepsilon_1-\varepsilon_0)
(\varepsilon_s-\varepsilon_0)e^{-2qa}]^{-2}, \nonumber
\end{eqnarray}
where $W$ is the input SAW power scaled to the width of the sound
track.
\begin{figure}[ht]
\centerline{
\includegraphics[width=7cm]{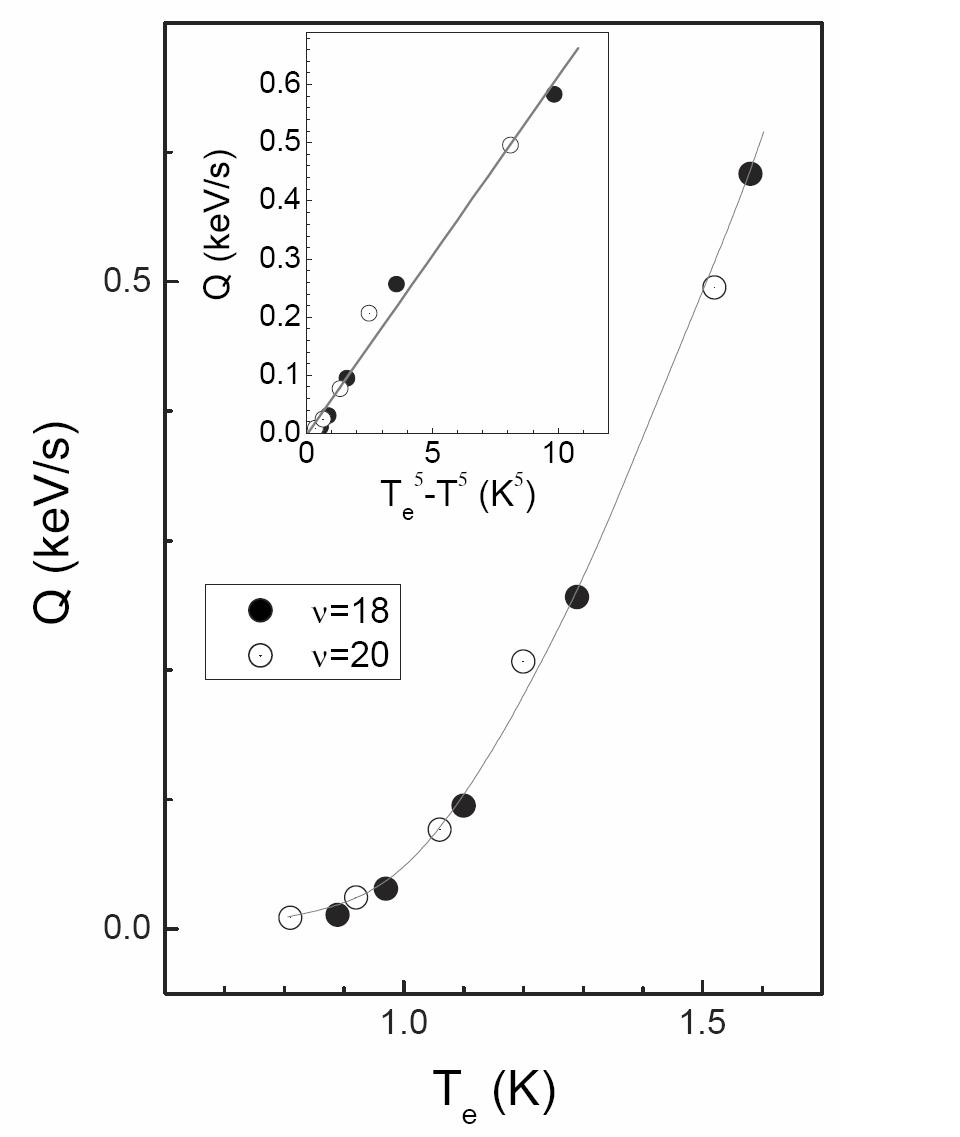}
} \caption{The energy losses rate per electron versus $T_e$; $T$=0.3 K,
$f$=30 MHz. The line is guide to the eye. Inset: dependence
of $Q$ on ($T_e^{\gamma}-T^{\gamma}$), $\gamma=5$. Line is the result of the
linear fit.
\label{QTehighNU}}
\end{figure}

The variation of $Q$ with $T_e$ obtained from the experimental
results is shown in  Fig.~\ref{QTehighNU} and could be well fitted
by the functional form $Q=A_5 (T_e^5-T^5)$, which corresponds to
energy relaxation via carrier scattering from the piezoelectric
potential of the acoustic phonons (PA-scattering) with strong
screening.

The value of $A_5$, found from the slope of the dependence $Q=A_5
(T_e^5-T^5)$, is $A_5\approx$ 62 eV/sK$^5$ (the theoretical
calculation using the piezoelectric constant $\beta_{14}=3.6 \times
10^4$ cgs gives $A_5 = $ 92 eV/sK$^5$)\cite{Karpus}. The value of
the energy relaxation time $\tau_{\varepsilon}$ can be determined
using the equation for arbitrary heating ~\cite{Gantmakher}:
\begin{eqnarray}
\tau_{\varepsilon}=\frac{\pi^2 k_B^2(T_e^2-T^2)}{6 E_F A_{5}
(T_e^5-T^5)}. \label{taue}
\end{eqnarray}
where $E_F$ is the Fermi energy. The energy relaxation time
calculated with known $A_5$ and Eq.~\ref{taue} depends on $T_e$, but
for low heating $\Delta T_e \ll T$ $\tau_{\varepsilon} \approx 5
\times 10^{-7}$ s and the condition $\omega \tau_{\varepsilon} \gg$1
is easily satisfied. This means that the energy relaxation time is
much greater than the period of the SAW oscillations and so the
heating processes depend on the time averaged SAW intensity.

Thus, relation (\ref{taurel}) is also satisfied which allows us to
explain the observed nonlinear effects in the delocalized electrons
region as simply the heating of free electrons by the electric field
of the SAW.

\subsubsection{Integer Quantum Hall Effect Regime} \label{QHE}

The presence of charged impurities leads to random fluctuations in
the electrostatic potential of the 2DEG and means that electrons
have to be excited to a percolation level for conduction to occur.
The influence of a strong static electric field on this activated
conductivity was considered theoretically in
Ref.~\onlinecite{Shklov} and shown to decrease the activation
energy, which can be interpreted as a lowering of the percolation
threshold. For the two dimensional case, the dependence of the
conductivity on electric field will look like:
\begin{eqnarray}
\sigma_1 = \sigma_1^0 \exp (\alpha E^{3/7} /k_B T),
\alpha = (C el_{sp} V_0)^{3/7}. \label{E3x7}
\end{eqnarray}
where $\sigma_1^0$ is the conductivity in the linear regime, $l_{sp}$ is
the spatial scale of the potential variation, which for modulation doped heterostructures could
be counted as equal to the spacer width, and $V_0$ is the random
fluctuation potential amplitude.
\begin{figure}[ht]
\centerline{
\includegraphics[width=8cm]{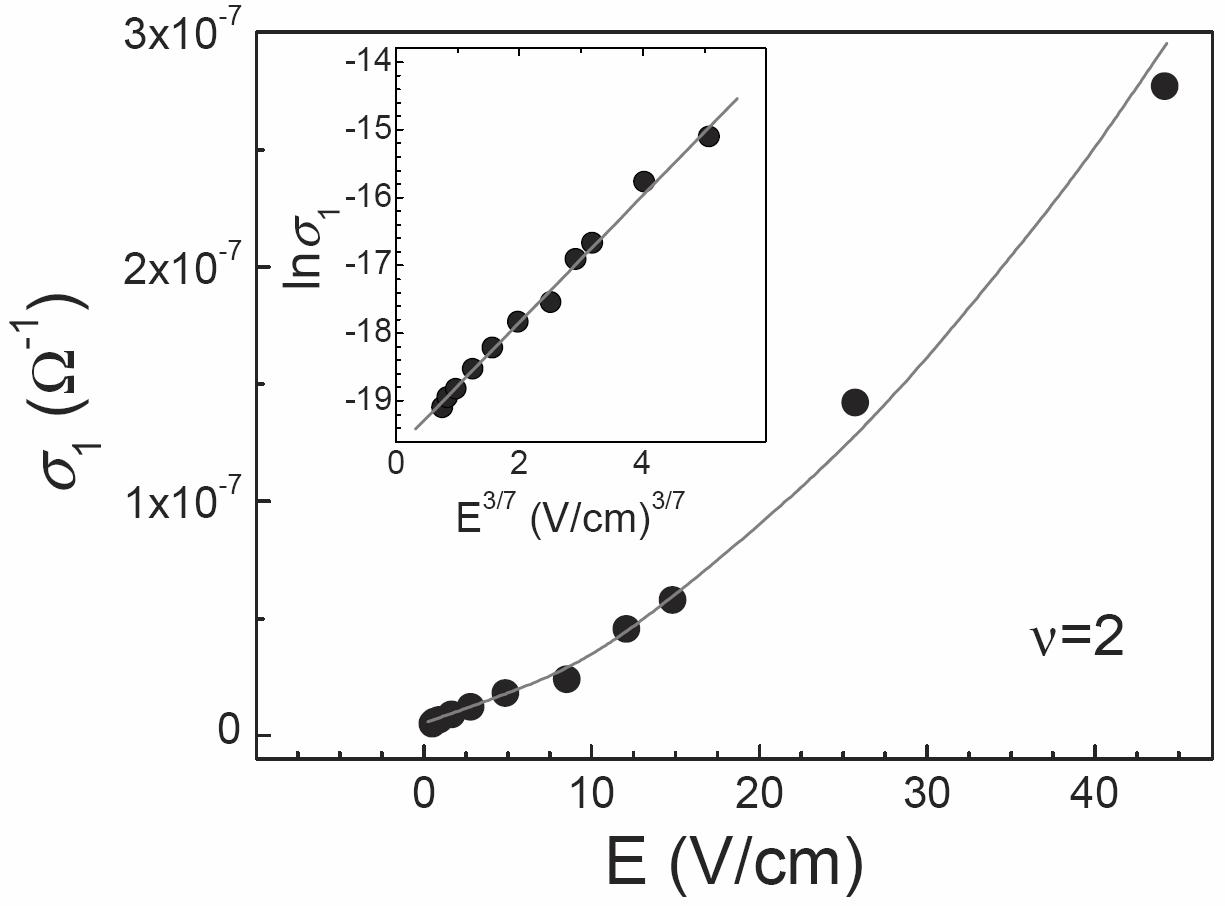}
} \caption{Variation of the real component of the ac conductivity
with the SAW electric field for $\nu$=2; $f$=30 MHz, $T$=0.3 K.
Inset: $\ln \sigma_1$ vs $E^{3/7}$. Line
is a result of the linear fit.\label{QTeQHE}}
\end{figure}

In our experiment, when the condition $\omega \tau_p \ll$ 1 is
valid, the wave (SAW) can be considered as stiffened, so to
interprete the nonlinear effects in the ac conductivity one may to
apply a theory obtained for the strong static electric field.
Although, we studied nonlinear effects at $T$ = 0.3 K, where the
conductivity no longer has an activation character, but was very
weakly dependent on temperature (as expected for the case of
two-site hopping), the dependence of the real conductivity
$\sigma_1$ on the electric field of surface acoustic wave shown in
Figure~\ref{QTeQHE} is nevertheless well described by the dependence
$\ln \sigma_1 \propto E^{3/7}$ predicted by equation~(\ref{E3x7}).
One can assume that the nonlinearity mechanism is complex: simply
heating the electrons in an electric field of the SAW leads to the
activated dependence of conductivity on temperature, the
nonlinearity in this case is characterized by nonlinear conduction
at the percolation level, as in Ref.~\onlinecite{Shklov}.

\subsubsection{Fractional Quantum Hall Effect Regime} \label{FQHEn}
\begin{figure}[ht]
\centerline{
\includegraphics[width=8cm]{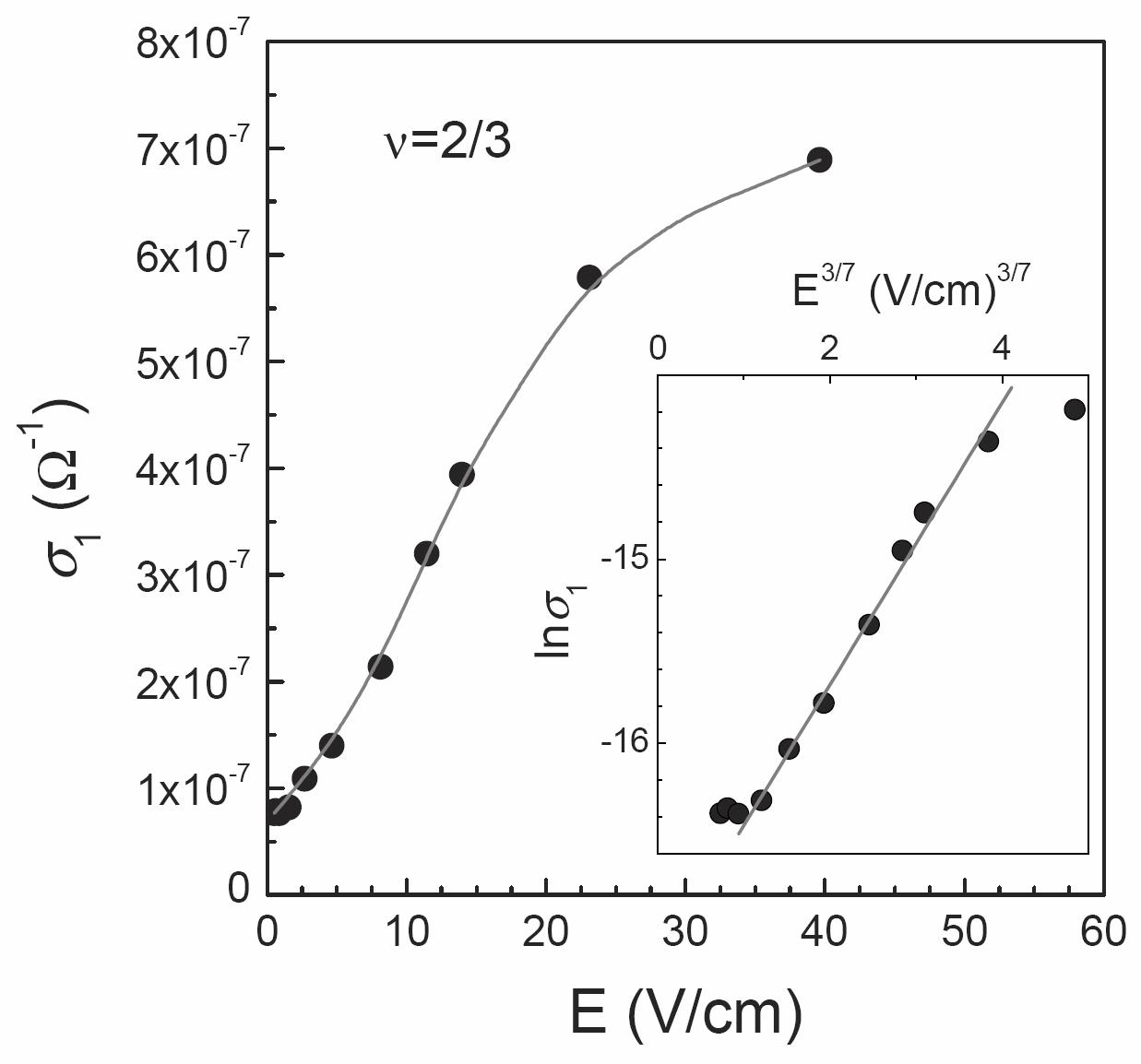}
} \caption{Dependence of the real component of the ac conductivity  on the SAW electric field $E$.
Inset: $\ln \sigma_1$ on $E^{3/7}$; $\nu$=2/3. Line
is a result of the linear fit.\label{Nonl2x3}}
\end{figure}

All that was said about the nonlinear effects in $\sigma_1$ in the
previous section for  $\nu$=2, is fully applicable to $\sigma_1$ in
the minimum of oscillations at $\nu$=2/3, but with a much larger
reason because activated conductivity for this case remains
practically down to $T$=0.3 K. Indeed Fig.~\ref{Nonl2x3} very
clearly demonstrates that the $\ln \sigma_1 \propto E^{3/7}$ law is
valid for the experiment at  $\nu$=2/3.

\subsection{Conductivity in a tilted magnetic field} \label{Tilted}

It is well known that in the fractional quantum Hall effect regime
at $\nu$=2/3 the spin polarized-unpolarized phase transition occurs
with changes of concentration or magnetic
field.~\cite{Eisen,2x3f,2x3h,Clark} Keeping in mind the description
of the FQHE many-body ground state as a one-electron state of
composite fermions, the $\nu$=2/3 FQHE state translates into the
$\nu$=2 IQHE state  of composite fermions. In this case, the spin
transition occurs due to crossing of CF LLs with different spin
direction as the magnetic field or concentration changes. This
transition has been investigated in numerous papers and by different
experimental methods. As the concentration of our sample was
$n=2\times 10^{11}$ cm$^{-2}$ we could not explore this transition
directly, since the $\nu$=2/3 state in our experiment was at $B$=12
T, while the spin transition occurs at lower magnetic fields. So we
could merely investigate the fully spin-polarized state. The usual
routine in this case is to study the activation energy of
conductivity in a tilted magnetic field, which for $\nu$=2/3 in the
fully spin-polarized state has been done using various
methods.~\cite{Eisen,2x3f,2x3h,Clark,Schulze,Haug,Boebinger,Khrapai}
However, the functional form reported for this dependence varies
from $\Delta E \propto B^1$
(Ref.~\onlinecite{Schulze},~\onlinecite{Khrapai}) to a sublinear one
with a saturation.~\cite{Eisen,Haug}
\begin{figure}[ht]
\centerline{
\includegraphics[width=8.5cm]{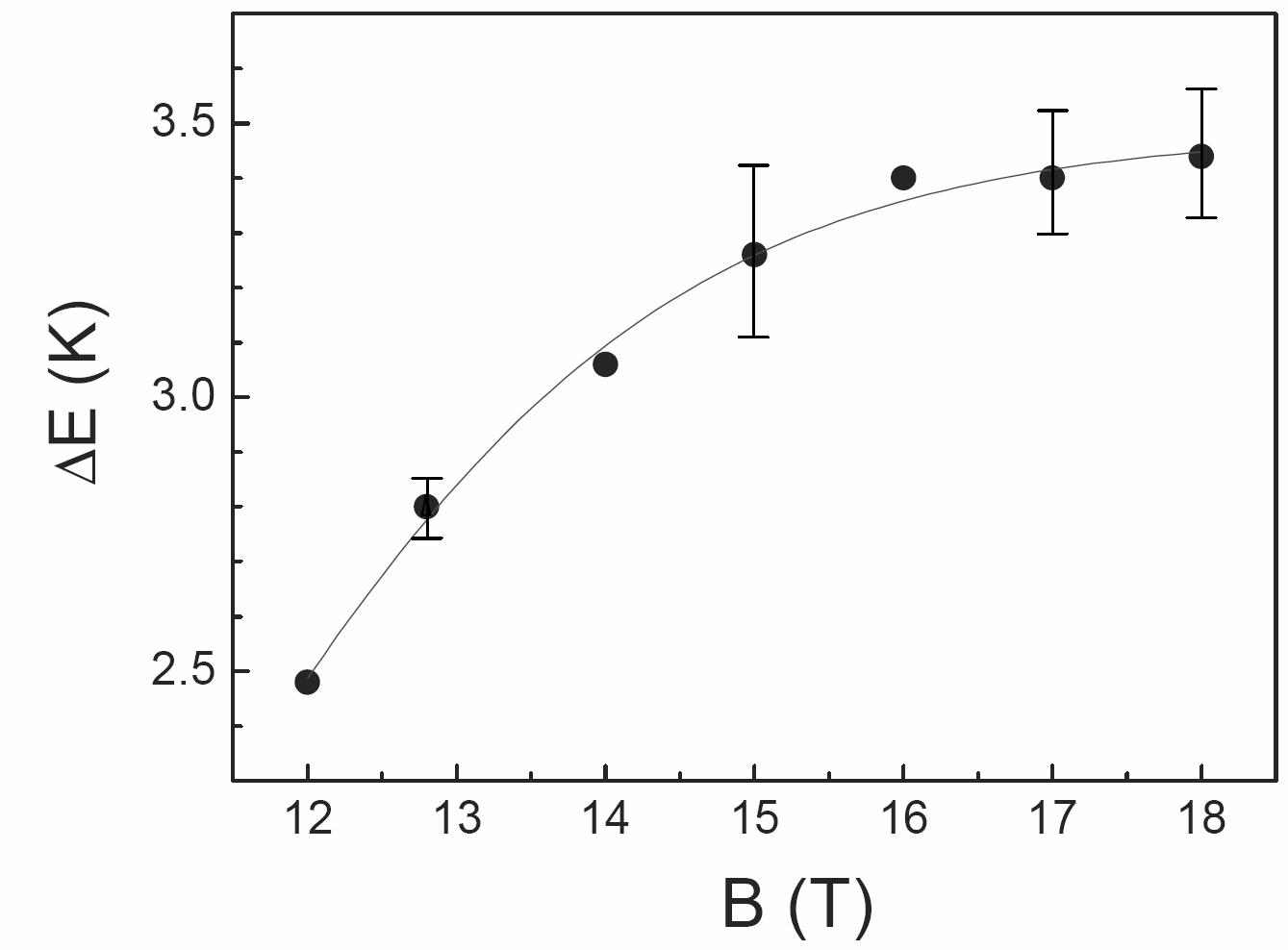}
} \caption{Dependence of the activation energy
$\Delta E$ on magnetic field for $\nu$=2/3; $f$=30 MHz.
The line is guide to the eye.\label{DETilt}}
\end{figure}

Tilting the magnetic field relative to a normal to the sample
surface enabled us to change the position of the conductivity
minimum at  $\nu$=2/3 from $B$=12.0 T up to $B$=17.8 T. We measured
the temperature dependence of   $\Gamma$ and $\Delta v/v$ in the
range of 0.4 K to 1.6 K for each tilt angle and calculated
$\sigma_1$ using formulas (1) and (2). The activation energy
$\Delta E$ was derived by constructing the Arrhenius plot $\ln
\sigma_1$ against 1/$T$ and the resulting dependence of $\Delta E$
on magnetic field is illustrated in Fig.~\ref{DETilt}.

It has been predicted, that in the FQHE regime the energy difference
between Landau levels  $\Delta E \propto B^{1/2}$. However, as in
all the previously cited works, our data shows that such dependence
is not obeyed. The reason for this is firstly that the method of
determining $\Delta E$ from  the temperature dependence of
conductivity minima leads one to determine an activation energy to
the mobility edge: the value of which  will differ from the value of
the gap in the Landau levels spectrum by the width of the region of
delocalized states. In order to obtain the actual gap between LL
centers one can instead apply the Liftshitz-Koesevich (LK) formula
to the damping of the SdH oscillations with temperature.~\cite{14a,14b}. In
Ref.~\onlinecite{14b} it was found that the energy gap between
centres of the spin polarised CF Landau levels does scale with the
effective magnetic field seen by the composite fermions and in Ref.
16 it was shown that changing the effective g-factor through
application of hydrostatic pressure has no effect on the gap of a
polarized state at 2/3. Secondly, tilting the field also changes the
in-plane component of magnetic field which serves to compress the
wavefunction closer to the interface, as modeled in
Ref.~\onlinecite{Gee}, and leads to deviations from the simple
square root behavior.

Nevertheless, it can be seen that acoustic techniques provide a
contact-free method of obtaining the information more usually sought
from dc transport experiments and are probing the similar physics,
with the same limitations. In principle, the LK formula could also
be applied to the acoustic results, but we are reluctant to do so
with the current data set since the oscillations are far from
sinusoidal.

\section{Conclusion} \label{Conclus}

In this paper we have demonstrated a method to determine the complex
high frequency conductivity in an $n$-GaAs/AlGaAs heterostructure
using acoustic contactless techniques in a "hybrid" geometry.

This work has shown that, in a structure exhibiting the quantum Hall
effect, one can separate the regions of delocalized and localized
electrons through application of a magnetic field.

At low magnetic fields, $B<$ 1 T, electrons are delocalized thus one
can determine the general parameters of the two-dimensional electron
gas using contactless acoustic methods: concentration, mobility,
Dingle temperature, and the transport, quantum and energy relaxation
times. The method is powerful and clearly justified since the
parameters determined by acoustic methods coincide to within 10$\%$
of those measured by direct current.~\cite{GeSiOur} At higher
fields, the mechanisms of low temperature high frequency
conductivity have been analyzed in the IQHE oscillations minima, and
shown to be predominately by hopping, which is well explained by the
"two-site" model.

In the fractional quantum Hall effect regime at $\nu$=2/3 the
conductivity has an activation character. The dependence of the
activation energy on magnetic field has been determined and shown to
follow a similar trend to measurements by other techniques.

A major feature of using this acoustic technique is that all the
conductivity measurements can be made without needing to fabricate
contacts. We expect this to be most desirable in evaluating emerging
novel material systems where the fabrication technology and surface
interactions are less well understood.

\subsection{Acknowledgments}

The authors are grateful to Yu.M. Galperin for useful discussions
and thank E. Palm, T. Murphy, J.H. Park, and G. Jones for help with
the experiments. The authors also thank J.J. Harris and C.T. Foxon
for growing the sample G157 back in 1986 and see this work as
excellent evidence of the longevity of their GaAs material. This
work was supported by grant of RFBR 11-02-00223, a grant of the
Presidium of the Russian Academy of Science, the Program
"Spintronika" of Branch of Physical Sciences of RAS. The NHMFL is
supported by the NSF through Cooperative Agreement No. DMR-0654118,
the State of Florida, and the DOE.

\vfill\eject

\end{document}